\documentstyle[preprint,tighten,aps,epsfig]{revtex}

\newcommand{\tr}{\mbox{Tr}\,}

\begin{document}
%\begin{titlepage}
%\begin{center}
%\vspace*{-1cm}
%\hfill
\preprint{LA PLATA--TH--01/02}

\title{\Large \bf Radiative decays of mesons in the NJL model}
\author{L.\ Epele, H.\ Fanchiotti, D.\ G\'omez Dumm\thanks{Fellow of
CONICET} and A.\ G.\ Grunfeld}
\address{\hfill \\Instituto de F\'{\i}sica La Plata, Depto.\ de
F\'{\i}sica, Fac.\ de Cs.\ Exactas, UNLP\\
C.C. 67, 1900 La Plata, Argentina} \maketitle
\begin{abstract}
We revisit the theoretical predictions for anomalous radiative decays
of pseudoscalar and vector mesons. Our analysis is performed in the
framework of the Nambu--Jona-Lasinio model, introducing adequate
parameters to account for the breakdown of chiral symmetry. The
results are comparable with those obtained in previous approaches.
\end{abstract}

%\vfill
%PACS numbers:
%\end{center}
%\end{titlepage}
%\newpage
%\pagestyle{plain}

\section{Introduction}

The analysis of vector and axial-vector meson physics represents a
nontrivial task from the theoretical point of view. This is
basically due to the characteristic energy scales at which these
particles manifest, namely an intermediate range between low
energy hadron physics and the high energy region where QCD can be
treated perturvatively. In order to deal with the subject, one
possible approach is to start from an effective meson Lagrangian;
another possible way is to derive the corresponding effective
interactions from QCD-inspired fermionic schemes.

In the last two decades, the development of chiral effective
models \cite{leutw} has provided a profitable framework to analyze
various phenomena related to the physics of vector and
axial-vector mesons \cite{Kay1,Kay2,Toni}. These models have been
built taking into account the symmetries of QCD, together with the
so--called axial anomaly introduced through the Wess-Zumino-Witten
(WZW) \cite{WZW} effective action. As an alternative approach, the
analysis of spin one meson physics can be performed by starting
with the Lagrangian proposed by Nambu and Jona--Lasinio (NJL)
\cite{NJL}. This Lagrangian is based in 4-fermion interactions,
showing the chiral symmetry of QCD, and leads to an effective
theory for scalar, pseudoscalar, vector and axial--vector mesons
after proper bosonization. The effective meson Lagrangian can be
obtained by constructing a generating functional, and evaluating
the real and imaginary parts of the fermion determinant
\cite{Rein1}.

In this work we follow this second approach, studying the
phenomenology of spin 1 mesons in the framework of the NJL
model. We propose a fermionic Lagrangian that shows an approximate
$U(3)_L\otimes U(3)_R$ chiral symmetry, explicitly violated
by current quark masses, as is the case of QCD in the
limit of large number of colours $N_C$. In particular, we revisit
the predictions given by this model for anomalous radiative decays
of vector mesons, taking into account the explicit breakdown of the
$SU(3)$ flavour symmetry to $SU(2)$ isospin symmetry with the
assumption $m_u = m_d \ne m_s$. We perform the bosonization of the NJL
theory by carrying out an expansion of the fermion determinant in terms
of the meson fields. This gives rise to a set of one--loop Feynman
diagrams \cite{Eguchi,Volkov1,Volkov2,Volkov3,Volkov4} from which
one can derive the relevant effective meson interactions. Taking
into account present experimental data, we analyze possible
input parameters for the model, in order to obtain an
acceptable phenomenological pattern for the observed anomalous
decays of pseudoscalar and vector mesons.

The paper is organized as follows: In Sect.\ II we introduce the
NJL Lagrangian and the bosonization technique. In Sect.\ III we
derive the effective Lagrangian to account for vector and
axial-vector meson interactions, considering the explicit
breakdown of flavour $SU(3)$ symmetry. This leads to some
relations between vector meson masses and decay constants that can
be written in terms of symmetry breaking parameters. Then, in
Sect.\ IV, we concentrate on the anomalous radiative decays of pseudoscalar
and vector mesons, which proceed through the $U(1)$ axial anomaly.
In Sect.\ V we discuss the numerical results for the corresponding
branching ratios, in comparison with the present experimental
information. Finally in Sect.\ VI we present our conclusions. The
Appendix includes a brief description of the regularization scheme
used throughout our analysis.

\section{Nambu--Jona--Lasinio Lagrangian and bosonization}

Chiral effective models have proved to reproduce the low
energy hadron phenomena with significant success. For that reason,
a significant effort has been done in order to derive these effective
schemes from a microscopic theory as QCD. The effective four
fermion Lagrangian proposed by Nambu and Jona--Lasinio \cite{NJL},
which shows the chiral symmetry of QCD, represents one
of the most popular ways to achieve this goal. The NJL model
provides, after proper bosonization, not only the expected
effective couplings for scalar, pseudoscalar, vector and
axial--vector mesons but also the Wess--Zumino--Witten
sector \cite{WZW} which accounts for the anomalous meson decays.
The explicit form of the NJL Lagrangian reads
\begin{eqnarray}
{\cal L} &=& \bar q (i \not\!\partial - \hat m_0)q + 2 G_1
\left[ (\bar q \frac{1}{2}\lambda^a q)^2 + (\bar q i \gamma_5
\frac{1}{2}\lambda^a q)^2 \right] \nonumber \\ & & - 2 G_2 \left[
(\bar q \gamma^\mu \frac{1}{2}\lambda^a q)^2 + (\bar q
\gamma^\mu \gamma_5\frac{1}{2} \lambda^a q)^2 \right],
\label{00}
\end{eqnarray}
where $q$ denotes the $N$-flavour quark spinor, $\lambda^a$,
$a=0,\dots,N^2-1$ are the generators of the $U(N)$ flavour group
(we normalize $\lambda^0=\sqrt{2/N}\,\openone$) and $\hat m_0$
stands for the current quark mass matrix, which explicitly breaks
the chiral symmetry. The coupling constants $G_1$ and $G_2$, as
well as the quark masses, are introduced as free parameters of the
model. In the absence of the mass term, the NJL Lagrangian shows
at the quantum level the $SU(N)_A \otimes SU(N)_V \otimes U(1)_V$
symmetry characteristic of massless QCD.

It is possible to reduce the fermionic degrees of freedom to
bosonic ones by standard bosonization techniques. With the
introduction of colorless boson fields and an appropriate use of
the Stratonovich identity, one obtains an effective action where
the fermions couple to the bosons in a bilinear form. The
couplings driven by $G_1$ in (\ref{00}) lead to the introduction
of scalar and pseudoscalar boson fields, whereas those carrying
$G_2$ lead to the inclusion of vector and axial--vector mesons.

To be definite, let us consider the vector meson sector with $N=3$
flavours, $u$, $d$ and $s$. By means of the Stratonovich identity,
the vector--vector coupling in (\ref{00}) can be transformed as
\begin{equation}
- 2 G_2(\bar{q} \gamma^\mu \frac{1}{2}\lambda^a q)^2 \rightarrow
-\frac{1}{4 G_2} \tr V_\mu^2 + i \bar{q}\gamma^\mu V_\mu q,
\label{masa}
\end{equation}
where $V_\mu \equiv -i \displaystyle \sum_{a=0}^8
V_\mu^a\;\lambda^a/2$. The spin 1 fields $V_\mu^a$ can be
identified with the usual nonet of vector mesons,
\begin{equation}
V = \frac{(-i)}{\sqrt{2}}\left(\begin{array}{ccc} \displaystyle
\frac{\rho^0}{\sqrt{2}} + \frac{\omega_8}{\sqrt{6}} +
\frac{\omega_1}{\sqrt{3}} &\rho^+ &K^{*+}\\ \rho^- &\displaystyle
-\frac{\rho^0}{\sqrt{2}} + \frac{\omega_8}{\sqrt{6}} +
\frac{\omega_1}{\sqrt{3}} &K^{*0}\\ K^{*-} &{\bar{K}}^{*0}
&\displaystyle - \frac{2\,\omega_8}{\sqrt{6}} +
\frac{\omega_1}{\sqrt{3}}
\end{array} \right)\,
\label{campos}
\end{equation}
which transform in such a way to preserve the
chiral symmetry of the original NJL Lagrangian (and therefore that
of QCD). Notice that the first term in the right hand side of Eq.\
(\ref{masa}) is nothing but a mass term for the vector fields
$V_\mu^a$, thus the vector--meson masses are governed by the
coupling $G_2$ in the NJL Lagrangian. It can be seen that these
masses are degenerate in the limit where the quark masses are
degenerate.

Now, the quark fields can be integrated out, leading to an
effective Lagrangian which only contains bosonic degrees of
freedom. This procedure can be carried out by taking into account
the generating functional
\begin{equation}
{\cal Z} = {\cal N} \int {\cal D}V {\cal D}q {\cal D}{\bar
q}\;\exp \left\{i\int dx^4 \left[-\frac{1}{4 G_2} \tr V_\mu^2 +
{\bar q} (i \not\!\partial - {\hat m_0} + i \not\!V) q
\right] \right\}
\label{genfunct}
\end{equation}
and performing the calculation of the fermion determinant (a
detailed analysis can be found in \cite{Rein1}). A similar
procedure can be followed for the full NJL Lagrangian (\ref{00}),
leading to the interactions involving scalar, pseudoscalar and
axial--vector bosons. In this way, the final effective Lagrangian
is written only in terms of spin 0 and spin 1 colorless hadron
fields.

Here we have performed the bosonization by carrying out an
expansion of the fermion determinant, which gives rise to a set of
one--loop Feynman diagrams \cite{Eguchi}. In order to build up the
final effective meson Lagrangian, we have obtained the local part
of the relevant interactions by taking the leading order of a
gradient expansion in powers of the external momenta, and considering
the dominant contributions in $1/N_C$. The divergent integrals have
been treated using a proper--time regularization scheme with a
momentum cut--off, keeping the leading contributions. The procedure
is developed in detail in the next section.

\section{Effective meson couplings and symmetry-breaking parameters}

In this section we describe the effective meson interactions derived
from the NJL Lagrangian. We begin by calculating the relevant one--loop
self--energy diagrams contributing to the meson kinetic and mass
terms, and those leading to the effective weak and strong meson decay
couplings. Notice that the meson fields introduced through the
bosonization technique acquire their dynamics via the self--energy
one--loop diagrams.

In our framework, the poles of the quark propagators
occur at the so--called constituent masses, i.e.\
dynamically generated masses which arise as a consequence of the
spontaneous breakdown of the chiral symmetry. The constituent masses
appear as solutions of Schwinger-Dyson self-consistency
equations (gap equations), which are governed by the scalar
coupling $G_1$ in the NJL Lagrangian (\ref{00}).
Here we will not concentrate in the scalar sector of the model,
thus the constituent quark masses, as well as the pseudoscalar
boson masses, will be considered as free parameters.

Our main interest is focused in the anomalous radiative decays of
vector mesons. Nevertheless, we need to consider also the axial--vector
sector of the model, since the axial--vectors mix with the
pseudoscalars at the one--loop level ($P-A$ mixing).

\subsection{Kinetic terms and masses of the spin-1 mesons}

The generating functional (\ref{genfunct}) gives rise to effective
kinetic terms for the spin-1 vector mesons via one--loop diagrams,
as shown in Fig.\ 1(a). The analysis in the case of the
axial--vector mesons can be performed in an analogous way, with
the additional ingredient of $P-A$ mixing given by Fig.\ 1(b).

The loop in Fig.\ 1(a) gives a contribution to the vector meson
self-energy given by
\begin{equation}
i N_C \int \frac{d^4 k}{(2\pi)^4}
\tr \frac{\not\!k - \not\!p + m_1}{(k-p)^2 - m_1^2}
\lambda^a \gamma_\mu \frac{\not\!k + m_2}{k^2 - m_2^2}\lambda^b \gamma_\nu\;,
\label{pimunu}
\end{equation}
where $m_1$ and $m_2$ are the constituent masses
of the quarks entering the loop, $N_C$ is the number of colors, and the
trace acts over the flavour and Dirac indices.

As stated above, we will take only the leading order in the external
momentum $p$, which means to evaluate the integral at $p=0$ after
extracting the relevant kinematical factors. In this case, this is
equivalent to consider only the divergent
piece of (\ref{pimunu}):
\begin{equation}
\Pi_{\mu\nu}^{(V)} = I_2(m_1, m_2)\,
\left[\frac{1}{3}\,(p_\mu p_\nu - p^2 g_{\mu\nu}) +
\frac{1}{2}\, (m_2 - m_1)^2\right]\,,
\label{pimunudiv}
\end{equation}
where
\begin{equation}
I_2(m_i, m_j) \equiv -i\,\frac{N_C}{(2 \pi)^4} \int d^4k
\frac{1}{(k^2 - m_i^2)(k^2 - m_j^2)}\,.
\end{equation}
As it was previously mentioned,
in order to regularize the divergence we use the
proper--time regularization scheme \cite{Ramon} with a cut--off
$\Lambda$, which will be treated as a free parameter of the model.
Details of the procedure can be found in the Appendix. We obtain
\begin{equation}
I_2(m_i, m_j) = \frac{N_C}{16 \pi^2}\int_0^1 dx\;
\Gamma\left(0,\frac{(m_i^2 - m_j^2) x + m_j^2 }{\Lambda^2}\right)\,.
\label{idos}
\end{equation}

{}From (\ref{pimunudiv}), the kinetic terms for the vector mesons in
the effective Lagrangian are given by
\begin{eqnarray}
{\cal L}_{kin}^{(V)} & = & -\frac{1}{4}\,\frac{2}{3}\, I_2(m_u, m_u)
\left[\rho_{\mu\nu} \rho^{\mu\nu} + 2
\rho^+_{\mu\nu} {\rho^-}^{\mu\nu} + \omega_{\mu\nu}
\omega^{\mu\nu} + \alpha\, \phi_{\mu\nu} \phi^{\mu\nu} \nonumber
\right.
\\ & & \left. + 2 \beta \left( {K^\ast}^+_{\mu\nu} {K^\ast}^{-\mu\nu} +
{K^\ast}^0_{\mu\nu} \bar K^{\ast 0\mu\nu} \right)\right] \, ,
\label{lkin}
\end{eqnarray}
where $V^{\mu\nu} \equiv \partial^\mu V^\nu - \partial^\nu V^\mu$, and
\begin{equation}
\alpha = \frac{I_2(m_s, m_s)}{I_2(m_u, m_u)} \;,
\qquad\qquad
\beta = \frac{I_2(m_u, m_s)}{I_2(m_u, m_u)}
\end{equation}
parameterize the magnitude of the $SU(3)$ flavour symmetry breaking.

The kinetic Lagrangian in (\ref{lkin}) has been expressed in terms of the
vector fields in (\ref{campos}), with the additional rotation
\begin{eqnarray}
\omega_8 & = & \phi \cos\theta_0 + \omega \sin\theta_0 \nonumber\\
\omega_1 & = & -\phi \sin\theta_0 + \omega \cos\theta_0 \;\;,
\end{eqnarray}
which diagonalizes the neutral sector. It is easy to see that
here the rotation is ``ideal'', i.e., the spin 1 mass
eigenstates $\rho$ and $\omega$ are composed by pure light $u$ and
$d$ quarks, while the $\phi$ meson is a bound state $\bar s s$. The
$\omega-\phi$ rotation angle is given by $\sin\theta_0 = 1/\sqrt{3}$.
Slight deviations from the ideal mixing condition will be considered
below to allow for the decay $\phi\to\pi^0\gamma$ observed recently.

The mass terms for the vector mesons are given by the $(V_\mu)^2$
term in (\ref{masa}), plus a divergent one--loop contribution
given by the second term in the square brackets in
(\ref{pimunudiv}), which vanishes in the $SU(3)$ flavour limit.
This leads to
\begin{eqnarray}
{\cal L}_{mass}^{(V)} & = & \frac{1}{8 G_2}\left[\rho_\mu \rho^\mu
+ 2 \rho^+_\mu {\rho^-}^\mu + \omega_\mu \omega^\mu + \phi_\mu
\phi^\mu + 2 {K^\ast}^+_\mu {K^\ast}^{-\mu} + 2 {K^\ast}^0_\mu
\bar K^{\ast 0\mu}\right] \nonumber \\
& & + (m_s - m_u)^2 \beta\, I_2(m_u,m_u)
({K^\ast}^+_\mu {K^\ast}^{-\mu} + {K^\ast}^0_\mu \bar K^{\ast
0\mu})\;.
\label{lvmass}
\end{eqnarray}
Notice that (as expected) the mass terms turn out to be diagonal
in the ($\omega,\phi$) basis.

We proceed now to the wave function renormalization required by the
kinetic terms in (\ref{lkin}). The vector
meson fields can be properly redefined by $V_\mu \to Z^{1/2}_V\, V_\mu$,
with
\begin{mathletters}
\begin{eqnarray}
Z_\rho^{-1} & = &  Z_\omega^{-1}  =
\frac{2}{3}\,I_2(m_u,m_u)
\label{vrena} \\
Z_{K^\ast}^{-1} & = &  \frac{2}{3}\,\beta
\,I_2(m_u,m_u) = \beta\, Z_\rho^{-1} \label{vrenb} \\
Z_\phi^{-1} & = &  \frac{2}{3}\,\alpha\,I_2(m_u,m_u)
= \alpha\, Z_\rho^{-1} \;. \label{vrenc}
\end{eqnarray}
 \label{vren}
\end{mathletters}
Then from (\ref{lvmass}) one obtains
\begin{equation}
m_\rho^2 = m_\omega^2 = \frac{Z_\rho}{4 G_2}\;,\qquad
m_{K^*}^2 = \frac{m_\rho^2}{\beta} + \frac{3}{2}(m_s - m_u)^2\;,\qquad
m_\phi^2 = \frac{m_\rho^2}{\alpha}\;,
\label{vmass}
\end{equation}
thus the $\phi$ meson mass can be written in terms of the $\rho$
mass and the flavour symmetry breaking parameter $\alpha$. In the
case of the $K^\ast$, the corresponding mass relation includes
both the parameter $\beta$ and a quark--mass dependent
contribution that arises from the loop in Fig.\ 1 (a).

A similar analysis can be performed for the axial--vector meson
sector. By replacing $\gamma_\alpha\to\gamma_\alpha\gamma_5$ in
the quark--meson vertices in Fig.\ 1 (a), one finds
\begin{equation}
\Pi_{\mu\nu}^{(A)} = \frac{2}{3}\,I_2(m_1, m_2)\,
\left[(p_\mu p_\nu - p^2 g_{\mu\nu}) - \frac{3}{2}\,
\frac{(m_2 + m_1)^2}{2}\right]\,,
\end{equation}
leading to the kinetic Lagrangian
\begin{eqnarray}
{\cal L}_{kin}^{(A)} & = & - \frac{1}{4}\,\frac{2}{3}I_2(m_u, m_u)
\left[{a_1}_{\mu\nu} a_1^{\mu\nu} + 2 {a_1}_{\mu\nu}^+
a_1^{-\mu\nu} + f_{1\mu\nu} f_1^{\mu\nu} \nonumber \right. \\ & &
\left. + 2 \beta {K^+_1}_{\mu\nu} K_1^{-\mu\nu} + 2 \beta
K^{0}_{1\mu\nu} {\bar{K}}_1^{0\mu\nu}
 + \alpha f'_{1\mu\nu} {f'}_1^{\mu\nu} \right]
\label{lkina}
\end{eqnarray}
and the mass terms
\begin{eqnarray}
{\cal L}_{mass}^{(A)} & = & \frac{1}{8 G_2}\left[{a_1}_\mu a_1^\mu
+ 2 {a_1^+}_\mu a_1^{-\mu} + f_{1\mu} f_1^\mu + f'_{1\mu}
{f'}_1^\mu + 2 {K^+_1}_\mu K_1^{-\mu} + 2 K^{0}_{1\mu}
{\bar{K}}_1^{0\mu}\right] \nonumber\\ & & + I_2(m_u,m_u) \left[ 2
m_u^2 ({a_1}_\mu a_1^\mu + 2 {a_1^+}_\mu a_1^{-\mu} + f_{1\mu}
f_1^\mu) \right. \nonumber \\ & & \left. + (m_u + m_s)^2 \beta
({K_1}_\mu^+ K_1^{-\mu} + {K_1}_\mu^0 \bar K_1^{0\mu}) + 2 m_s^2
\alpha f'_{1\mu} {f'}_1^\mu \right] \, . \label{amlag}
\end{eqnarray}
As in the vector meson case, the Lagrangian is diagonal in the
chosen basis, where $f_1$ and ${f'}_1$ are obtained from the
$U(3)$ states $A_8$ and $A_1$ through an ideal rotation. Notice
that both neutral and charged axial--vector boson masses receive a
contribution proportional to $G_2^{-1}$, which arises from field
transformations as in Eq.\ (\ref{masa}), plus additional
(positive) contributions proportional to the quark masses. In this
way, the axial--vector mesons are in general expected to be
heavier than their vector meson counterparts.

Finally, we have to take into account the mixing between the
axial--vector mesons and the pseudoscalars. This requires the
analysis of the pseudoscalar--axial vector couplings and
the pseudoscalar kinetic terms, given by the
one--loop diagrams in Fig.\ 1 (b) and (c) respectively.
The effective couplings read
\begin{eqnarray}
{\cal L}_{mix} & = & -i I_2(m_u,m_u) \left[2 m_u (\partial_\mu
\pi^0 a_1^\mu +
\partial_\mu \pi^+ a_1^{-\mu} +
\partial_\mu \pi^- a_1^{+\mu} + \partial_\mu \eta_u
f^{\mu})\nonumber \right. \\
& & + \left.\beta (m_u + m_s)
(\partial_\mu K^+ K_1^{\mu-} +
\partial_\mu K^- K_1^{\mu+} + \partial_\mu K^0 \bar{K}_1^{0\mu}+
\partial_\mu \bar{K}^0 K_1^{0\mu})\nonumber \right.\\
& & + \left. 2 \alpha m_s
\partial_\mu \eta_s f'^\mu\right]\, ,
\end{eqnarray}
while the pseudoscalar kinetic Lagrangian is given by
\begin{eqnarray}
{\cal L}_{kin}^{(P)}& = & -\frac{1}{2} I_2(m_u,m_u)
\left[\partial_\mu \pi^0 \partial^\mu \pi^0 + 2
\partial_\mu \pi^+ \partial^\mu \pi^-
+ \partial_\mu \eta_u \partial^\mu \eta_u + \alpha\,
\partial_\mu \eta_s \partial^\mu \eta_s \nonumber \right. \\
& & + \left. 2 \beta\, \partial_\mu K^+ \partial^\mu K^-
+ 2 \beta\, \partial_\mu K^0 \partial^\mu \bar{K}^0\right]\,.
\end{eqnarray}

Once again, we have chosen a basis for the neutral fields in which
the states $\eta_u$, $\eta_s$ are obtained from the $U(3)$ states
$\eta_8$, $\eta_1$ through an ideal rotation. However, these states
cannot be treated as approximate
mass eigenstates due to the presence of the $U(1)_A$ anomaly, which
breaks the $U(3)$ symmetry down to $SU(3)$. Although formally of
order $1/N_C$, this anomaly leads to a relatively large mass splitting
between the observed $\eta$ and $\eta'$ physical states. Still,
large $N_C$ considerations are shown to be powerful enough to deal with
the interactions of the $\eta_1$ and $\eta_8$ fields as members of an
$U(3)$ nonet. As stated, we will not concentrate here in the pseudoscalar
mass sector, and consequently, in the $U(1)_A$ symmetry breaking mechanism
responsible for the $\eta'$ mass. In the NJL framework, one way to proceed
is to include the so-called 't~Hooft interaction \cite{thooft,kleva},
which emulates the effect of the anomaly.
Instead of following this way (which means to include six-fermion
interaction vertices) we will introduce the $\eta-\eta'$ mixing
angle as a parameter of the model. Thus we write $\eta_u$ and
$\eta_s$ in terms of the physical states as
\begin{eqnarray}
\eta_u & = & \cos \varphi_P\;\eta + \sin \varphi_P \; \eta' \nonumber\\
\eta_s & = & -\sin \varphi_P\;\eta + \cos \varphi_P\; \eta'\;.
\label{mix}
\end{eqnarray}
In order to write the effective Lagrangian in terms of the
physical fields, one needs not only the relevant wave function
renormalizations but also the diagonalization of the $P-A$
couplings. This can be achieved by means of the transformations
\begin{equation}
P \to Z^{1/2}_P P\;,
\qquad\qquad
A_\mu \to Z^{1/2}_A\, A_\mu + C_P Z_P^{1/2}\,\partial_\mu P\;,
\end{equation}
where $P=\pi$, $K$, $\eta_u$, $\eta_s$, and
\begin{eqnarray}
C_\pi & = & i\,2\, I_2(m_u,m_u) Z_\rho \frac{m_u}{m_{a_1}^2}\;
=\;C_{\eta_u} \nonumber \\
C_K & = & i\,2\, I_2(m_u,m_u) Z_\rho
\frac{(m_u+m_s)}{2\,m_{K_1}^2} \nonumber \\
C_{\eta_s} & = & i\,2\,
I_2(m_u,m_u) Z_\rho \frac{m_s}{m_{f'}^2}\;. \label{betas}
\end{eqnarray}
The wave function renormalization for the pseudoscalar sector yields
\begin{eqnarray}
Z_\pi^{-1} & = & Z_{\eta_u}^{-1} = I_2(m_u,m_u) \left(1 - \frac{6
m_u^2}{m^2_{a_1}}\right)\nonumber \\ Z_K^{-1} & = & \beta
I_2(m_u,m_u) \left[1 - \frac{3}{2}\frac{(m_u + m_s)^2
}{m^2_{K_1}}\right] \nonumber \\ Z_{\eta_s}^{-1} & = & \alpha
I_2(m_u,m_u) \left(1 - \frac{6 m_s^2}{m^2_{f'}}\right),
\label{zetas}
\end{eqnarray}
where the axial--vector meson masses can be read from
(\ref{amlag}). As can be immediately seen from (\ref{lkina}), the
wave function renormalization for the axial--vector mesons is
equivalent to that performed in the vector meson sector, c.f.\
Eq.\ (\ref{vren}). Using these relations and those in
(\ref{vmass}), the axial--vector meson masses turn out to be
constrained by
\begin{eqnarray}
m_{a_1}^2 & = & m_\rho^2 + 6 m_u^2 \nonumber \\ m_{K_1}^2 & = &
m_{K^*}^2 + 6 m_u m_s
\label{amass} \\
m_{f'}^2 & = & m_\phi^2 + 6 m_s^2
\nonumber\;.
\end{eqnarray}

\subsection{Weak and strong decay constants}

We want to analyze the ability of the effective model
under consideration to describe the low energy behaviour of
pseudoscalar and vector mesons. Let us begin by studying some
fundamental decay processes, in order to evaluate the relevant
decay constants in terms of the model parameters.

The pseudoscalar weak decay constants $F_\pi$ and $F_K$ can be
obtained from the one--loop transitions represented in Fig.\ 2,
where $P=\pi$, $K$. As in the self--energy case, these diagrams
are found to be logarithmically divergent and can be regularized
using the proper--time scheme. We will keep as before only the
leading contributions in the limit of vanishing external momenta,
which amounts to take only the divergent piece of the loop
integrals.

The decay constants $F_P$ are defined as usual by
\begin{equation}
\langle 0|J_{\mu 5}(0)|P(q)\rangle = i\,\sqrt{2}\, F_P\, q_\mu \;.
\end{equation}
It is easy to see that the contributions from Fig.\ 2 (a) and (b)
lead to
\begin{eqnarray}
F_\pi & = & 2\, I_2(m_u,m_u) m_u Z_\pi^{1/2} \left(1 +
i\, 2\, m_u C_\pi\right) \nonumber \\
& = & 2\, I_2(m_u,m_u) m_u Z_\pi^{1/2} \left( 1 - 6
\frac{m_u^2}{m^2_{a_1}}\right) \nonumber \\
F_K & = & \rule{0cm}{.5cm}
I_2(m_u,m_u)\,\beta\,(m_u+m_s)\, Z_K^{1/2}
\left[1 + i\,(m_u+m_s) C_K\right] \nonumber \\
& = & \rule{0cm}{.8cm}
I_2(m_u,m_u) \beta\, (m_u + m_s) Z_K^{1/2}
\left[1 - \frac{3}{2} \frac{(m_u + m_s)^2}{m^2_{K_1}}\right]\;,
\end{eqnarray}
where have made use of Eqs.\ (\ref{betas}). In these equations,
the first terms arise from the graph in Fig.\ 2 (a), whereas the
contributions proportional to $C_{\pi,K}$ are due to the
$P-A$ mixing diagrams. Using now Eqs.\ (\ref{zetas}), we
end up with the simple relations
\begin{equation}
F_\pi = \frac{2\,m_u}{Z_\pi^{1/2}}\;,\qquad\qquad
F_K = \frac{(m_u + m_s)}{Z_K^{1/2}}\;.
\label{fpika}
\end{equation}

Next, let us study the basic decay constants $g_\rho$,
$g_{K^\ast}$ and $g_\phi$, which account for the
strong decays of the vector mesons into two pseudoscalars.
The decay constants can be defined from the total decay
rates according to
\begin{eqnarray}
\Gamma(\rho\to \pi\pi)& = & \frac{g_\rho^2}{4\pi}\, \frac{m_\rho}{12}
\left(1-\frac{4m_\pi^2}{m_\rho^2}\right)^{3/2} \nonumber \\
\Gamma(K^\ast\to K\pi) & = & \rule{0cm}{.8cm}
\frac{g_{K^\ast}^2}{4\pi}\, \frac{m_{K^\ast}}{16}
\left\{\left[1-\frac{(m_K+m_\pi)^2}{m_{K^\ast}^2}\right]
\left[1-\frac{(m_K-m_\pi)^2}{m_{K^\ast}^2}\right]\right\}^{3/2} \nonumber \\
\Gamma(\phi\to K K) & = & \rule{0cm}{.5cm}
\Gamma(\phi\to K^+K^-) + \Gamma(\phi\to K^0\bar K^0)
\nonumber \\
& = & \rule{0cm}{.8cm}
\frac{g_\phi^2}{4\pi}\, \frac{m_\phi}{24} \left[
\left(1-\frac{4 m_{K^+}^2}{m_\phi^2}\right)^{3/2}+
\left(1-\frac{4 m_{K^0}^2}{m_\phi^2}\right)^{3/2}\right]\,.
\label{strongdec}
\end{eqnarray}
Notice that for the $\phi\to K K$ width we have taken into account the
mass difference between the neutral and charged
kaons. Though tiny, this mass difference becomes important in view
of the very narrow phase space allowed.

Within the effective model analyzed here, the decay rates
(\ref{strongdec}) are dominated by the one--loop graphs
shown in Fig.\ 3. As before, we keep here the dominant,
logarithmically divergent contribution, and the leading
order in the external momenta (the corresponding diagrams
with two axial--vector legs vanish in this limit). From
the evaluation of the $\rho\to\pi\pi$ and $\rho\to
a_1\pi$ diagrams, and using
(\ref{vren}), (\ref{betas}) and (\ref{zetas}), we find
\begin{equation}
g_\rho = I_2(m_u,m_u) Z_\rho^{1/2} Z_\pi \left( 1 +
i\,2\,m_u\, C_\pi\right) = Z_\rho^{1/2}\;.
\label{grho}
\end{equation}
In the case of $g_\phi$ the evaluation of the dominant
contribution is more involved, since the loop includes two
different quark propagators. The loop integral has the form
\begin{equation}
I_3 \equiv -i\,\frac{N_C}{(2 \pi)^4} \int d^4k
\frac{k^2}{(k^2 - m_u^2)(k^2 - m_s^2)^2}\,. \label{int3}
\end{equation}
Once again the divergence is logarithmic, thus it is natural to
express the integral in terms of $I_2(m_i,m_j)$. However, owing to
the flavour symmetry breakdown, there is an ambiguity at
the moment of choosing the infinite part of (\ref{int3}). Here we
follow the criterion of respecting the natural hierarchy given by
the quark content of the decaying vector meson, that means, we use
the $k^2$ factor in the integrand to remove the $m_u^2$ pole.
We have then
\begin{equation}
I_3 \to \alpha\, I_2(m_u,m_u) \;.
\end{equation}
Taking into account the wave function renormalization for the
$\phi$ and $K$ mesons, together with the corresponding
pseudoscalar--axial vector mixing, we end up with
\begin{equation}
g_\phi = \frac{\sqrt{\alpha}}{\beta}\,  g_\rho\;. \label{gf}
\end{equation}

In the same way we proceed for the case of the $K^\ast$ to two
meson decay. Here, due to the $SU(3)$ symmetry breaking,
the analogous of Eq.\ (\ref{grho}) for $g_{K^\ast}$ leads to a
relation between $g_{K^\ast}$ and $g_\rho$ which can be written
in terms of $Z_{\pi,K}$ and $C_{\pi,K}$. At first approximation,
however, the dependence on these parameters cancels and one
can write
\begin{equation}
g_{K^\ast} \simeq g_\rho \;.
\label{gk}
\end{equation}

In this way, both $\phi$ and $K^\ast$ strong decay constants can
be predicted in terms of the measured value of $g_\rho$ and the
symmetry breaking parameters $\alpha$ and $\beta$. The corresponding
numerical evaluation will be done in Sect.\ V.

\section{Anomalous radiative decays of pseudoscalar and vector mesons}

Our goal is to test the ability of this relatively simple,
NJL--like model to reproduce the observed pattern for the
anomalous radiative decays of pseudoscalar and vector mesons once
the breakdown of chiral symmetry is taken into account. These
decays are mainly originated by the $U(1)_A$ anomaly. In the
framework of our model, they occur through the well--known
triangle graphs shown in Fig.\ 4, which lead to an anomalous
Wess--Zumino--Witten effective Lagrangian. The corresponding
diagrams including an axial--vector which mixes with the
pseudoscalar meson gives a vanishing contribution \cite{waka}.
This can be seen {\em e.g.} by looking at the effective anomalous
Lagrangian obtained by gauging the Wess--Zumino--Witten action, as
shown in Ref.\ \cite{Kay2}.

The couplings between the vector mesons and the outgoing
(on--shell) photons follow from the assumption of vector meson
dominance of the electromagnetic interactions. In our framework,
these couplings can be obtained from quark--loop diagrams
connecting photons and vector mesons (see Ref.\
\cite{Rein1,Volkov3} for details). After the proper inclusion of
the symmetry breaking parameters $\alpha$ and $\beta$, the
$V-\gamma$ couplings read
\begin{equation}
j_\mu^{em} = e\, m_\rho^2 \,Z_\rho^{-1/2}\left(\rho_\mu +
\frac{1}{3}\; \omega_\mu + \frac{\sqrt{2}}{3}
\frac{m_\phi^2}{m_\rho^2} \sqrt{\alpha}\; \phi_\mu \right)\;.
\end{equation}

Let us begin by analyzing the decays of
the neutral pseudoscalars $\pi^0$, $\eta$ and $\eta'$ into two
photons. These processes arise at the one--loop level from the
triangle diagrams shown in Fig.\ 4 (a). The diagrams are found
to be convergent, and the limit of vanishing external momenta
can be taken trivially. For $\pi^0\to\gamma\gamma$ we get
\begin{equation}
\Gamma(\pi^0\to\gamma\gamma) =
\frac{\alpha^2 m_\pi^3}{64\pi^3}
\left(\frac{Z_\pi^{1/2}}{2 m_u}\right)^2 =
\frac{\alpha^2 m_\pi^3}{64\pi^3 F_\pi^2} \;,
\label{pi2g}
\end{equation}
where $\alpha$ is the electromagnetic fine structure constant.
This expression coincides with the result obtained in standard
Chiral Perturbation Theory \cite{leutw}. In the case of $\eta$ and
$\eta'$ decays to two photons, one has the additional
problem of $\eta-\eta'$ mixing. In our framework, instead of
dealing with the $U(3)$ states $\eta_1$ and $\eta_8$, it is
natural to work with the $\eta_u$ and $\eta_s$ states defined
above, which determine the flavour content of the quark propagators
in the loop. Taking into account the corresponding pseudoscalar
wave--function renormalizations, we find
\begin{eqnarray}
\Gamma(\eta\to\gamma\gamma) & = &
\frac{\alpha^2 m_\eta^3}{64\pi^3}
\left[\frac{5}{3 F_\pi}\cos\varphi_P -
\frac{\sqrt{2}}{3 F_s}\sin\varphi_P\right]^2 \nonumber \\
\Gamma(\eta'\to\gamma\gamma) & = &
\frac{\alpha^2 m_{\eta'}^3}{64\pi^3}
\left[\frac{5}{3 F_\pi}\sin\varphi_P +
\frac{\sqrt{2}}{3 F_s}\cos\varphi_P\right]^2 \;,
\end{eqnarray}
where, in analogy with $F_\pi$ and $F_K$, we have defined
\begin{equation}
F_s \equiv \frac{2m_s}{Z_{\eta_s}^{1/2}}\;.
\label{fs}
\end{equation}

Now we analyze the radiative vector meson decays $V\to P\gamma$.
The corresponding widths can be conveniently parameterized as
\begin{equation}
\Gamma(V \rightarrow P \gamma) = \frac{\alpha g_{\rho}^2}{24\pi}
\,\frac{C^2_{VP}}{F_\pi^2}
\left(\frac{M^2_V - M^2_P}{4 \pi M_V}\right)^3\;,
\label{rate}
\end{equation}
where the coefficients $C_{VP}$ can be obtained from
the loop diagrams shown in Fig.\ 4 (b). Clearly,
the evaluation of the relevant triangle quark loops is entirely
similar to that performed for the $P\to\gamma\gamma$ case.

For the processes $\rho^0,\omega\to \pi^0\gamma$ we can use
Eqs.\ (\ref{vrena}), (\ref{fpika}) and (\ref{grho}) to eliminate
the pseudoscalar and spin 1 wave function renormalization factors.
In this way the coefficients $C_{\rho^0\pi^0}$ and $C_{\omega\pi^0}$
are simply given by
\begin{equation}
C_{\rho^0\pi^0}= 1\;,\quad\quad\quad C_{\omega\pi^0}=3\;.
\end{equation}
The same procedure applies in the case of $V\to\eta,\eta'\gamma$
processes, with the additional inclusion of the $\eta-\eta'$
mixing angle and the flavour symmetry breaking parameter $\alpha$.
We have
\begin{equation}
\begin{array}{lcl}
C_{\rho^0\eta} &=& 3 \cos \varphi_P \\
C_{\omega\eta} &=&  \cos \varphi_P \\
C_{\phi\eta} &=& 2 F_\pi \sin \varphi_P/(\sqrt{\alpha} F_s) \\
C_{\phi\eta'} &=& 2 F_\pi \cos \varphi_P/(\sqrt{\alpha} F_s)\;,
\end{array}
\end{equation}
where $\varphi_P$ is the $\eta_u-\eta_s$ mixing angle defined in
(\ref{mix}).

For the decays $K^\ast\to K\gamma$, the calculation is slightly
more complicated due to the presence of different constituent masses
for the quark propagators in the loop. The coefficients depend now
on the symmetry breaking parameter $\beta$ and the mass
ratio $\lambda\equiv m_s/m_u$ according to
\begin{equation}
\begin{array}{lcl}
C_{K^{\ast +}K^+} &=& f(\lambda)F_\pi/(\sqrt{\beta} F_K) \\
C_{K^{\ast 0}K^0} &=& g(\lambda)F_\pi/(\sqrt{\beta} F_K)\;,
\end{array}
\end{equation}
where functions $f(\lambda)$ and $g(\lambda)$ are defined as
\begin{eqnarray}
f(\lambda) & = & \left[\frac{1 + 6\lambda - \lambda^2}{2(1 - \lambda^2)} +
\frac{\lambda\, (2\,\lambda^2+1)\ln \lambda^2}{(1-\lambda^2)^2} \right]
\nonumber \\
g(\lambda) & = & \left[1 + \frac{\lambda\ln \lambda^2}{(\lambda^2 -1)}
\right].
\end{eqnarray}

Finally let us consider the decays $\eta'\to\rho\gamma$ and
$\eta'\to\omega\gamma$, which can be treated in a completely
similar way just taking into account the spin 0 character of the
decaying particle when averaging over the initial spin states.
Using the same parameterization as in (\ref{rate}) we obtain
\begin{equation}
\begin{array}{lcl}
C_{\rho^0\eta'} & = & 3 \sin \varphi_P \\
C_{\omega\eta'} & = & \sin \varphi_P \;.
\end{array}
\end{equation}

We have skipped in this discussion the decay $\phi\to\pi^0\gamma$
since the corresponding coefficient $C_{\phi\pi}$ vanishes in our
framework, owing to the purely strange flavour content of the $\phi$
meson. However, in view of the present experimental results, it
is interesting to remove the ideal mixing angle condition in the
$\omega-\phi$ sector allowing for a tiny nonstrange component for
the $\phi$ field. If this is parameterized by a small mixing
angle $\varphi_V$, one trivially gets \cite{Volkov4}
\begin{equation}
C_{\phi\pi^0} = C_{\omega\pi^0}\,\sin\varphi_V = 3\,\sin\varphi_V\;.
\end{equation}
The present experimental value for the branching ratio
$\phi\to\pi^0\gamma$ leads to a mixing angle $\varphi_V$ of about
$3.2^\circ$ \cite{Volkov4,bramon}, which does not represent a
significant change in the other, nonvanishing radiative decay widths
involving $\phi$ and $\omega$.

\section{Model parameters and numerical analysis}

In order to perform the phenomenological analysis of the model,
and to obtain definite predictions for pseudoscalar and vector meson
decay processes, it is necessary to establish a strategy, choosing a
suitable set of parameters to be used as input values.

Let us begin with the basic decay processes analyzed in Section III.
Using the $\pi^\pm$ and $K^\pm$ weak decays, it is possible to
write the pseudoscalar wave function renormalization factors
$Z_\pi$ and $Z_K$ in terms of the well--measured decay constants
$F_\pi$ and $F_K$, which will be taken as input parameters. In
the same way, the wave function renormalization factor $Z_\rho$
can be obtained from the phenomenological value $g_\rho\simeq 6$
arising from (\ref{strongdec}). Now, from (\ref{idos}) and
(\ref{vrena}), we can relate $g_\rho$ with the loop integral
$I_2(m_u,m_u)$ to estimate
\begin{equation}
\frac{m_u}{\Lambda} \simeq 0.26 \;.
\end{equation}
Next we use the mass relations (\ref{vmass}) to determine a
phenomenological acceptable set of values for the cut--off
$\Lambda$ and the strange quark mass $m_s$. A reasonable choice is
\begin{equation}
m_s\simeq 510 \mbox{ MeV}\;, \qquad\qquad\qquad
\Lambda\simeq 1.2 \mbox{ GeV}\;,
\end{equation}
which lead to $m_\phi=0.99$ GeV (exp.\ 1.02 GeV)
and $m_{K^\ast}=0.91$ GeV (exp.\ 0.89 GeV), with a light quark
mass $m_u=m_d\simeq 310$ MeV. The symmetry breaking
parameters $\alpha$, $\beta$ and $\lambda$ defined in the previous
sections are then given by
\begin{equation}
\alpha\simeq 0.6\;, \qquad\qquad\qquad
\beta\simeq 0.76\;, \qquad\qquad\qquad
\lambda\simeq 1.6\;.
\label{symbr}
\end{equation}

Using the relations (\ref{gf}) and (\ref{gk}), the phenomenological
value of $g_\rho$, together with the results in (\ref{symbr}) allow
to predict the values for the strong decay constants $g_{K^\ast}$
and $g_\phi$. In this way we obtain
\begin{equation}
\begin{array}{llr}
\Gamma(K^\ast \to K\pi) & \simeq & 44 \mbox{ MeV} \\
\Gamma(\phi \to K^0 \bar K^0) & \simeq & 1.3 \mbox{ MeV}
\\ \Gamma(\phi \to K^+ \bar K^-) & \simeq & 2.0 \mbox{ MeV}
\end{array}
\end{equation}
while the corresponding experimental values read
$\Gamma(K^\ast \to K\pi) = 50.8\pm .9$ MeV, $\Gamma(\phi \to K^0
\bar K^0) = 1.51\pm 0.03$ MeV and $\Gamma(\phi \to K^+ K^-) =
2.19\pm 0.03$ MeV \cite{tabla}. The results are in agreement
with the predicted values within an accuracy of a 15\% level,
which can be taken as a lower bound for the expected intrinsic
theoretical error of the model.

As an alternative procedure, we could have derived the values for
$F_\pi$ and $F_K$ from the relations (\ref{zetas}), where the
axial-vector meson masses can be taken either from Eqs.\ (24) or
just as input parameters. For instance, with the chosen value of
$m_u$ we can obtain from Eq.\ (\ref{amass}) a prediction for the
$a_1$ mass of about 1080 MeV \cite{tabla}. This value is somewhat
low, but once
again agrees within a 15\% accuracy with the present experimental
result of $m_{a_1}=1230\pm 40$ MeV. In fact, it is natural to
expect the model to describe the axial-vector meson sector only in
a roughly approximate way, due to the relatively high masses
involved, the broad character of the resonances and the admixtures
with close states with same quantum numbers. Hence we choose to
rely on the well-known low energy decay constants $F_\pi$ and $F_K$
as our input magnitudes.

Finally, using the analytical expressions obtained in the previous
section, we can look at the model predictions for pseudoscalar and
vector meson radiative decays. In the case of the decays involving
$\eta$ and $\eta'$ mesons, it is necessary to fix two more input
parameters, which can be chosen as the decay constant $F_s$ (or
equivalently the pseudoscalar wave function renormalization
$Z_{\eta_s}$) and the $\eta_1-\eta_8$ mixing angle $\theta_P$.
The latter is related with $\varphi_P$ by
\begin{equation}
\varphi_P = \frac{\pi}{2} - \theta_0 + \theta_P \;,
\end{equation}
where $\theta_0\simeq 35.3^\circ$ is the ``ideal'' rotation angle
introduced above.

Our numerical results are shown in Table I, where we quote the values
corresponding to four different sets for $F_s$ and $\theta_P$,
together with the present experimental widths \cite{tabla}. In order
to get the theoretical results, we have used the physical values for
the pseudoscalar and vector meson masses entering Eq.\ (\ref{rate}).
We have chosen the values of $F_s$ and $\theta_P$ that lead to a better
agreement for the whole set of decays, always having in mind that
an intrinsic theoretical error of at least 15\% can be expected from
the NJL scheme. In general, we find that the best results for
$\theta_P$ lie within the usually expected range between $-10^\circ$ and
$-20^\circ$ \cite{bramon,Holstein}, while the model seems to prefer a
rather large value for
the ratio $F_s/F_\pi$, higher than 3/2. We quote for completeness the
results corresponding to a ratio $F_s/F_\pi=1.4$, favored
by large-$N_C$ arguments \cite{feldmann}, which leads in this case
to too large widths for the decays involving $\eta$ and $\phi$ mesons.

When varying $F_s$ and $\theta_P$, and comparing with the experimental
values, it is found that an accurate agreement for certain widths worsens
the predictions in other cases. Nevertheless, considering the relatively
important intrinsic theoretical error of the model, it can be said that
the values in Table I show a phenomenologically acceptable pattern for
the pseudoscalar and vector meson anomalous radiative decays. Our results
are found to be of similar quality to those presented e.g.\ in
Ref.\ \cite{Kay2}, where the authors start from a nonlinear sigma
model, and Ref.\ \cite{Volkov4}, where a different
parameterization of the chiral--flavour symmetry breaking is
introduced (notice that several experimental values have changed
in the last years).

\section{Conclusions}

Starting from an effective four--fermion NJL Lagrangian, we have
analyzed the phenomenology associated with anomalous radiative
decays of pseudoscalar and vector mesons. The effective interactions
between spin 0 and spin 1 mesons have been derived from the
bosonization of the fermionic theory through the evaluation of
the relevant quark loop diagrams, taking the leading order
both in the external momenta and $1/N_C$ power expansions.
The divergent loops have been treated using a proper--time
regularization scheme with a momentum cut--off $\Lambda$.

We have taken into account the breakdown of the chiral symmetry
considering constituent quark masses $m_u=m_d\neq m_s$.
The departure of the effective couplings from the global $SU(3)$
flavour symmetry limit has been introduced through the
parameters $\alpha$ and $\beta$, which correspond to properly
regularized fermion loop integrals.

In order to perform the phenomenological analysis, we have chosen
a few fundamental input parameters, namely the basic decay
constants $F_\pi$, $F_K$ and $g_\rho$, the $\rho$ meson mass,
the momentum cut--off (or equivalently the light quark mass $m_u$),
and the constituent mass of the strange quark $m_s$. We have also
taken into account the physical values of the pseudoscalar meson
masses. To account for the decays
involving $\eta$ and $\eta'$ mesons, this set has been complemented
with the parameter $F_s$ and the $\eta-\eta'$ mixing angle $\theta_P$.
We have shown that with the choice $\Lambda\simeq 1.2$ GeV and
$m_s\simeq 510$ MeV this simple model leads to an acceptable
pattern for the vector meson mass spectrum, as well as the
branching ratios for the strong decays $\rho\to\pi\pi$, $\phi\to K
K$ and $K^\ast\to K\pi$. The agreement is found within a 15\% level,
that can be taken as a lower bound for the intrinsic theoretical
error of the model.

We have evaluated within this framework the branching ratios for
anomalous radiative decays of pseudoscalar and vector mesons.
In the case of those processes involving the $\eta$ and $\eta'$, the
contrast between the predictions of our model and the
present experimental data is optimized for a $\theta_P$ mixing
angle between $-10^\circ$ and $-20^\circ$, in agreement with
usual expectations. On the other hand, we find that the preferred
ratio $F_s/F_\pi$ lies in the range 1.75 to 2, which
is a rather high value in comparison with the result
arising from large $N_C$ considerations. In general, taking into
account the relatively large theoretical error,
we see that the model leads to reasonably good predictions for
the main radiative pseudoscalar and vector meson decay widths.
The quality of our results is comparable to that
obtained in previous works which use different effective models
and/or parameterizations for the flavour symmetry breaking
effects.

\appendix

\section*{}

The divergent integrals $I_2(m_i, m_j)$ have been regularized
within the proper--time scheme \cite{Ramon}. One
makes use of the relation
\begin{equation}
\frac{1}{A^{n+1}}\; \to\;
\frac{1}{n!}\int_{1/\Lambda^2}^\infty ds\;s^n e^{-sA}\;,
\end{equation}
which holds for $\Lambda$ sufficiently large.
If we let $\Lambda$ be a momentum cut--off for our theory,
we have for equal masses $m_i=m_j$
\begin{eqnarray}
& & I_2(m,m)\,  = \, \frac{N_C}{(2\pi)^4}
\int d^4k_E\;\frac{1}{(k_E^2 + m^2)^2} \nonumber \\
& & \to \; \frac{N_C}{(2\pi)^4}
\int d^4k_E \int_{1/\Lambda^2}^\infty
ds\; s\; e^{-s(k_E^2+m^2)} = \frac{N_C}{16\pi^2}\;
\Gamma\left(0,\frac{m^2}{\Lambda^2}\right)\,,
\label{idosapp}
\end{eqnarray}
where $k_E$ is the momentum in Euclidean space, and
$\Gamma(\alpha,x)$ is the Incomplete Gamma Function,
\begin{equation}
\Gamma(\alpha,x)\equiv \int_x^\infty\; t^{\alpha-1}\;e^{-t}\;dt\;.
\end{equation}

In the case where $m_i\neq m_j$, Eq.\ (\ref{idosapp}) can be
generalized using the Feynman parameterization
\begin{equation}
\frac{1}{(k_E^2 + m^2_i)(k_E^2 + m^2_j)} =
\int^1_0 dx\; \frac{1}{(k_E^2 + B^2(x))^2} \;,
\end{equation}
where $B^2(x) = m_i^2 + (m^2_j - m^2_i)\, x$. Using the
same regularization procedure as before one ends up with
\begin{equation}
I_2(m_i,m_j) \to \frac{N_C}{16\pi^2} \int_0^1 dx\;
\Gamma\left(0,\frac{B^2(x)}{\Lambda^2}\right)\;.
\end{equation}
%____________________________________________________________________________
\begin{figure}[ht]
\vspace{1cm}
    \begin{center}
       \setlength{\unitlength}{1truecm}
       \leavevmode
       \hbox{
       \epsfysize=3.4cm
       \epsffile{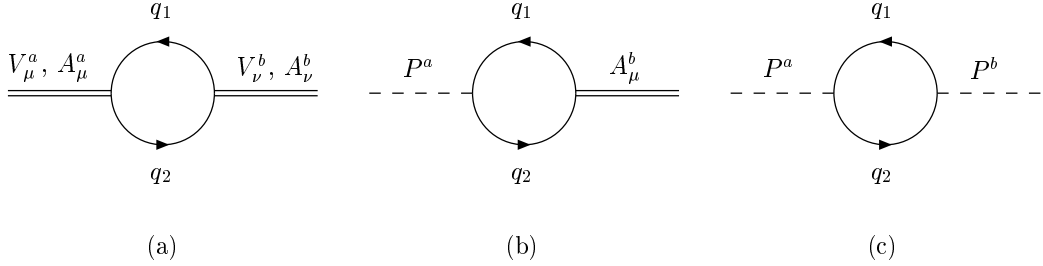}
       }
    \end{center}
    \caption{(a) One--loop self--energy diagrams for vector and axial vector
mesons, (b) pseudoscalar--axial vector mixing diagram, (c)
pseudoscalar self--energy diagram.}
    \protect\label{fig1}
\end{figure}
%____________________________________________________________________________

%___________________________________________________________________________
\begin{figure}[ht]
\vspace{1cm}
    \begin{center}
       \setlength{\unitlength}{1truecm}
       \leavevmode
       \hbox{
       \epsfysize=3.4cm
       \epsffile{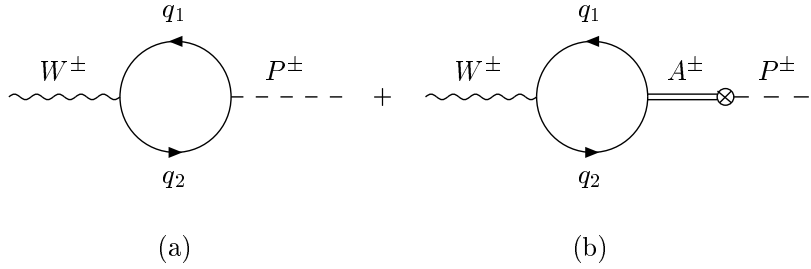}
       }
    \end{center}
    \caption{One--loop diagrams accounting for weak decays of pseudoscalar
    mesons.}
    \protect\label{fig2}
\end{figure}
%___________________________________________________________________________
%
%___________________________________________________________________________
\begin{figure}[ht]
\vspace{1cm}
    \begin{center}
       \setlength{\unitlength}{1truecm}
       \leavevmode
       \hbox{
       \epsfysize=3.9cm
       \epsffile{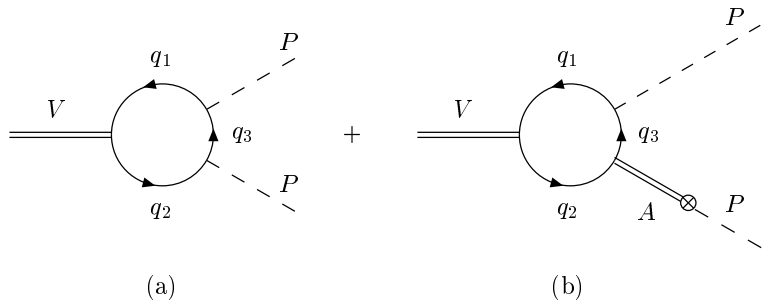}
       }
    \end{center}
    \caption{One--loop diagrams accounting for strong decays of vector mesons
    into two pseudoscalars.}
    \protect\label{fig3}
\end{figure}
%___________________________________________________________________________
%
%___________________________________________________________________________
\begin{figure}[ht]
\vspace{1cm}
    \begin{center}
       \setlength{\unitlength}{1truecm}
       \leavevmode
       \hbox{
       \epsfysize=3cm
       \epsffile{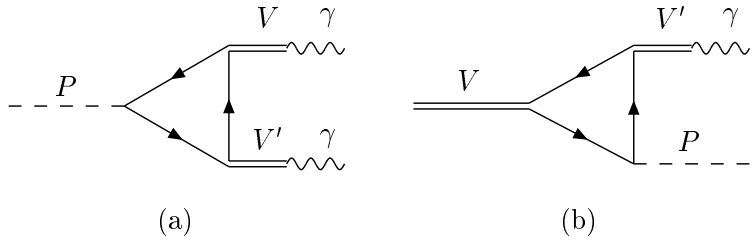}
       }
    \end{center}
    \caption{One--loop diagrams accounting for anomalous decays of
    (a) pseudoscalars and (b) vector mesons}
    \protect\label{fig4}
\end{figure}

%\newpage

\begin{table}
\caption{Radiative decays of pseudoscalar and vector mesons.
The second and third columns show the model predictions
for different values of the ratio $F_s/F_\pi$ and the $\eta_1-
\eta_8$ mixing angle $\theta_P$. The present experimental values
for the decay rates are quoted in the last column.}
\vspace{1cm}
%\hfill
\begin{tabular}{cccccc}
Process & \multicolumn{4}{c}{$\Gamma_{NJL}$ (KeV)} &
$\Gamma_{exp}$ (KeV) \\ \rule{0cm}{.7cm} & $F_s/F_\pi=2$ &
$F_s/F_\pi=1.75$ & $F_s/F_\pi=1.75$ & $F_s/F_\pi=1.4$ & \\
 & $\theta_P=-10^\circ$ & $\theta_P=-15^\circ$
 & $\theta_P=-20^\circ$ & $\theta_P=-20^\circ$ &
\\ \hline $\pi^0\rightarrow\gamma\gamma$ & $7.7\times 10^{-3}$ &
$7.7\times 10^{-3}$ & $7.7\times 10^{-3}$ & $7.7\times 10^{-3}$ &
$(7.7\pm 0.6)\times 10^{-3}$ \\
$\eta\rightarrow\gamma\gamma$ & 0.53 & 0.63 & 0.76 & 0.71
& $0.46\pm 0.04$ \\
$\eta'\rightarrow\gamma\gamma$ & 4.97 & 4.48 & 3.79 & 4.16
& $4.28\pm 0.44$ \\
$\eta'\rightarrow\rho^0\gamma$ & 108 & 89.5 & 71.1 & 71.1
& $59\pm 5$ \\
$\eta'\rightarrow\omega\gamma$ & 10.0 & 8.2 & 6.5 & 6.5
& $6.1 \pm 0.8$ \\
$\rho^0 \rightarrow \pi^0 \gamma$ & 85 & 85 & 85 & 85
& $102\pm 25$ \\
$\omega\rightarrow \pi^0 \gamma$ & 806 & 806 & 806 & 806
& $720\pm 40$ \\
$\rho^0\rightarrow \eta \gamma$ & 51 & 60 & 68 & 68
& $36\pm 14$ \\
$\omega\rightarrow \eta \gamma$ & 6.6 & 7.7 & 8.8 & 8.8
& $5.5\pm 0.8$ \\
%$\phi \rightarrow \pi^0 \gamma$ & 0 & 0 & $5.80\pm 0.58$ \\
$\phi \rightarrow \eta \gamma$ & 65.0 & 70.0 & 55.6 & 86.9
& $57.8\pm 1.5$ \\
$\phi \rightarrow \eta'\gamma$ & 0.30 & 0.46 & 0.52 & 0.82
& $0.30\pm 0.16$ \\
${K^\ast}^+ \rightarrow K^+ \gamma$ & 62 & 62 & 62 & 62
& $50\pm 5$ \\
${K^\ast}^0 \rightarrow K^0 \gamma$ & 160 & 160 & 160 & 160
& $117\pm 10$
\end{tabular}
\end{table}


\begin{references}

\bibitem{leutw} J. Gasser and H. Leutwyler, Ann. Phys. (N.Y.) {\bf 158},
142 (1984); Nucl. Phys. {\bf B250}, 465 (1985).

\bibitem{Kay1} \"O. Kaymakcalan, S. Rajeev and J. Scechter, Phys. Rev. D {\bf 30},
594 (1984); \"O. Kaymakcalan and J. Schechter, Phys. Rev. D {\bf
31}, 1109 (1985); J. Schechter, Phys. Rev. D {\bf 34}, 868 (1986).

\bibitem{Kay2} H. Gomm, \"O. Kaymakcalan and J. Schechter, Phys. Rev. D {\bf 30},
2345 (1984); M. Spali\'nski, Z. Phys. C {\bf 42}, 595 (1989).

\bibitem{Toni} G. Ecker, J. Gasser, A. Pich and E. de Rafael, Nucl.
Phys. {\bf B321}, 311 (1989).

\bibitem{WZW} J. Wess and B. Zumino, Phys. Lett. {\bf 37B}, 95
(1971).

E. Witten, Nucl. Phys. {\bf B223}, 422 (1983).

\bibitem{NJL} Y. Namb\'u and G. Jona-Lasinio, Phys. Rev. {\bf 122},
345 (1961); {\bf 124}, 246 (1961).

\bibitem{Rein1} D. Ebert and H. Reinhardt, Nucl. Phys. B {\bf 271}, 188
(1986); M.\ Wakamatsu and W.\ Weise, Z.\ Phys. A {\bf 331}, 173 (1988).


\bibitem{Eguchi} T. Eguchi, Phys. Rev. D {\bf 14}, 2755 (1976).

\bibitem{Volkov1}D. Ebert and M. Volkov, Z. Phys. C {\bf 16}, 205 (1983).

\bibitem{Volkov2}M. Volkov, Ann. Phys. {\bf 157}, 282 (1984).

\bibitem{Volkov3}D. Ebert, A. Ivanov and M. Volkov, Fortschr. Phys. {\bf 37}, 487 (1989).

\bibitem{Volkov4}M. Volkov, Phys. Part. Nucl. {\bf 24}, 35 (1993).

\bibitem{Ramon} M. Jaminon, R. M\'endez Gal\'ain, G. Ripka and P.
Strassart, Nucl. Phys. A {\bf 537}, 418 (1992).

\bibitem{thooft} G. 't Hooft, Phys. Rev. Lett. {\bf 37}, 8 (1976).

\bibitem{kleva} S. P. Klevansky, Rev. Mod. Phys. {\bf 64}, 649 (1992).

\bibitem{waka} M. Wakamatsu, Ann. Phys. {\bf 193}, 287 (1989).

\bibitem{bramon} A. Bramon, R. Escribano and M.
Scadron, Eur. Phys. J. C {\bf 7}, 271 (1999).

\bibitem{tabla} D. E. Groom {\em et al.}, Eur. Phys. J. C {\bf 15}, 1 (2000).

\bibitem{Holstein} E. P. Venugopalan and B. R. Holstein,
Phys. Rev. D {\bf 57}, 4397 (1998).

\bibitem{feldmann} T. Feldmann, Int. J. Mod. Phys. A {\bf 15}, 159 (2000).

\end{references}
\end{document}